\documentclass{aa}
\newcommand {\hii}{H\,{\sc ii}} 
\newcommand {\hi}{H\,{\sc i}} 
\newcommand {\kms}{\relax \ifmmode {\,\rm km\,s}^{-1}\else \,km\,s$^{-1}$\fi}
\newcommand {\ha}{H$\alpha$}

\newcommand {\n}{N\,11}
\newcommand {\x}{{\it XMM-Newton}}

\usepackage{graphicx}
\usepackage{txfonts}
\usepackage{rotating}
\usepackage{array}

\begin{document}

   \title{{\it XMM-Newton} observations of the giant \hii\ region N\,11 in the LMC\thanks{Based on observations collected with \x, an ESA Science Mission with instruments and contributions directly funded by ESA Member States and the USA (NASA).}}

   \subtitle{}

   \author{Y. Naz\'e\thanks{Research Fellow FNRS (Belgium)}\inst{1},
	  I.I. Antokhin\thanks{On leave from Sternberg Astronomical Institute, Moscow University, Russia}\inst{1,2},
          G. Rauw\thanks{Research Associate FNRS (Belgium)}\inst{1},
	  Y.-H. Chu \inst{3}, E. Gosset$^{\dagger}$\inst{1}, 
	  \and J.-M. Vreux \inst{1}}

   \offprints{Y. Naz\'e \email{naze@astro.ulg.ac.be}}

   \institute{Institut d'Astrophysique et de G\'eophysique;
	Universit\'e de Li\`ege;
	All\'ee du 6 Ao\^ut 17, Bat. B5c;
	B 4000 - Li\`ege;
	Belgium 
\and Astronomy and Astrophysics Group;
	Department of Physics and Astronomy;
	Kelvin Building, University of Glasgow;
	Glasgow G12 8QQ;
	United Kingdom
\and Astronomy Department;
University of Illinois at Urbana-Champaign;
	1002 West Green Street;
	 Urbana, IL 61801;
	USA
             }
\authorrunning{Naz\'e et al.}
   \date{Received 6 November 2003 / Accepted 21 January 2004}

   \abstract{
Using the sensitive {\it XMM-Newton} observatory, we have observed 
the giant \hii\ region \n\ in the LMC for $\sim$30~ks. We have 
detected several large areas of soft diffuse X-ray emission along with 37 
point sources. One of the most interesting results is the possible
association of a faint 
X-ray source with BSDL 188, a small extended object of uncertain nature. 

The OB associations in the field-of-view (LH9, LH10 and LH13) are all detected
with \x, but they appear very different from one another. 
The diffuse soft X-ray emission associated with LH9 peaks near 
HD 32228, a dense cluster of massive stars. The combined emission of 
all individual massive stars of LH9 and of the superbubble they have 
created is not sufficient to explain the high level of emission observed: 
hidden SNRs, colliding-wind binaries and the numerous pre-main sequence 
stars of the cluster are most likely the cause of this discrepancy. 
The superbubble may also be leaking some hot gas in the ISM since faint, 
soft emission can be observed to the south of the cluster. 
The X-ray emission from LH10 consists of three pointlike sources and a 
soft extended emission of low intensity. The two brightest point
sources are clearly associated with the fastest expanding bubbles 
blown by hot stars in the SW part of the cluster. The total X-ray emission 
from LH10 is rather soft, although it presents a higher temperature than the 
other soft emissions of the field. The discrepancy between
the combined emission of the stars and the observed luminosity 
is here less severe than for LH9 and could be explained in terms of hot 
gas filling the wind-blown bubbles. 
On the other hand, the case of LH13 is different: it does not harbour any 
extended emission and its X-ray emission could most probably be explained by 
the Sk $-66^{\circ}$41 cluster alone. 

Finally, our \x\ observation included simultaneous observations with 
the OM camera that provide us with unique UV photometry of more than 6000 
sources and enable the discovery of the UV emission from the SNR N11L.

   }

   \maketitle
\section{Introduction}

Massive stars are known to deeply influence the structure and dynamics 
of their environment. Their fast stellar winds combined with their huge 
mass-loss rates and their powerful 
explosion as supernovae (SNe) can shape the interstellar medium (ISM), 
creating various structures from small 
wind-blown bubbles around single stars to large superbubbles around 
OB associations. This collective action can best be understood by 
studying nearby \hii\ regions, with embedded OB associations 
containing hundreds of massive stars. In this context, we have chosen 
to study the \n\ complex (Henize \cite{hen}) in the Large Magellanic Cloud 
(LMC).

\n\ is the second largest \hii\ region in the LMC after 30 Doradus, 
and it may constitute a more evolved version of this latter nebula (Walborn 
\& Parker \cite{wal92}). It harbors several associations of massive stars: 
LH9, LH10, LH13
and LH14 (Lucke \& Hodge \cite{luc}). Its structure is complex and reflects the 
interactions between the stars and their environment. The central cluster, 
LH9, is surrounded by a filamentary shell of $\sim$120~pc in diameter. 
The combined action of stellar winds and supernova explosions from the 
members of the cluster has carved a hollow cavity in the surrounding 
ISM, thereby creating the shell which is also called a superbubble. 
The hot shocked winds/ejecta that fill this cavity emit X-rays (Mac 
Low et al. \cite{mac}) and provide the pressure to drive the superbubble 
shell expansion. Only supernovae outside the superbubble can 
produce distinct supernova remnants (SNRs) such as N11L at the western 
edge of the N11 complex (Williams et al. \cite{wil}). 

\begin{figure*}
\begin{center}
\end{center}
\caption{Three colour image of the \n\ region as seen by the combined EPIC 
cameras. The red, green and blue colours correspond respectively 
to the $0.4-1.0$ keV, $1.0-2.0$ keV and $2.0-10.0$ keV energy bands. 
Note that this image is based on a pn event list that was filtered 
using \#XMMEA\_EP and $0\le$ pattern $\le12$. This event list was only 
used for aesthetic reasons in the aim of creating Figs. \ref{color} 
\& \ref{totfield}, not for any scientific analysis.
\label{color}}
\end{figure*}

The action of LH9 on its surroundings has also triggered a burst of star 
formation at the periphery (Rosado et al. \cite{ros}), leading 
to the birth of the three other OB associations. Situated to the north 
of LH9, LH10 is still embedded in its natal cloud but its most massive 
components have already begun to blow bubbles around them (Naz\'e et 
al. \cite{naz}). The stellar population of LH9 and LH10 has been 
studied by Parker et al. (\cite{par}, hereafter PGMW) and Walborn 
et al. (\cite{wal99}). To the east of LH9 lies LH13, which appears 
to contain two tight clusters, Sk $-66^{\circ}$41 and HNT
(Heydari-Malayeri et al. \cite{hey00}). However, the ages and 
radial velocities indicate that the $\le$ 5 Myr old cluster 
Sk~$-66^{\circ}$41 is associated with the surrounding \hii\ region 
and the  $\sim$100 Myr old cluster HNT is unrelated and most probably just 
a line-of-sight object (Heydari-Malayeri et al. \cite{hey00}). 
Situated at the northeast outskirts of \n, LH14 is the least studied 
of the four OB  associations; the existence of a few massive stars in 
LH14 has been  illustrated in a photometric study by 
Heydari-Malayeri et al. (\cite{hey87}). 

\begin{figure*}
\begin{center}
\end{center}
\caption{Combined EPIC image of \n\ in the energy range $0.4-10$ keV. The 
detected sources are labelled. The image has been binned by a factor 
of 50, to obtain a pixel size of 2.5\arcsec.
\label{totfield}}
\end{figure*}

In summary, the \n\ complex harbours a variety of phenomena associated 
with massive stars (e.g. bubbles and  superbubbles), and even contains 
a SNR. All these objects 
should be associated with some hot gas. \n\ thus provides an interesting 
target for deep X-ray observations. A {\it ROSAT} investigation only permitted 
to detect some X-ray emission (Mac Low et al. \cite{mac}, Dunne et al. 
\cite{dun}), but the unprecedented sensitivity of {\it XMM-Newton} enables 
us to study this region in more details.

In the following sections, we will first describe the observations and 
the X-ray sources detected in the field. Next, we will focus on the main 
components of \n: the SNR, the superbubble, LH10 and LH13. Finally, we 
will conclude in Section~6.

\section{Observations}
\subsection{X-ray data}

\n\ was observed with {\it XMM-Newton} (Jansen et al. \cite{jan}) 
in the framework of the guaranteed time of the Optical Monitor 
Consortium during revolution 325, on 17 Sept. 2001, for approximately 
30 ks. The two EPIC MOS cameras (Turner et al. \cite{tur})
were used in full window mode, and the EPIC pn instrument (Str\"uder 
et al. \cite{str}) was operated in extended full 
frame mode. A thick filter was added to reject optical light.
These instruments are sensitive to radiation in the 0.4-10.0~keV range,
and the on-axis spatial resolution of \x\ is about 5\arcsec.

We used the Science Analysis System (SAS) software version 5.4.1. 
to reduce the EPIC data. These data were first processed through 
the pipeline chains, and then filtered.
For EPIC MOS, only events with a pattern between 0 and 12 and 
passing through the \#XMMEA\_EM filter were considered. For EPIC pn, 
we kept events with flag$=$0 and a pattern between 0 and 4. 
To check for contamination by low energy protons, we have further 
examined the light curve at high energies (Pulse Invariant channel number$>$10000, $\sim$E$>$10 keV, 
and with pattern$=$0). No background flares were detected. The resulting 
useful exposure times are 32.5~ks, 32.5~ks, and 25.4~ks for EPIC MOS1, 
MOS2 and pn, respectively. Further analysis was performed using the SAS 
version 5.4.1. and the FTOOLS tasks. The spectra were analysed and fitted 
within XSPEC v11.0.1. 

Figures \ref{color} and \ref{totfield} show images of the field from the combined EPIC data. Figure \ref{color} provides a three colour image, 
in which numerous point sources and diffuse, soft emissions are clearly 
seen. Figure \ref{totfield} gives the identification of the point 
sources detected in the field (see Sect. 3).  

\begin{figure}
\begin{center}
\end{center}
\caption{ Optical Monitor (OM) image of \n\ through the UV filter 
UVM2. Note that the FOV is 
smaller than the EPIC FOV. At the rightmost side of the image, the SNR 
N11L can be seen as a faint diffuse ring. Diffuse halos around bright 
stars are artefacts.
\label{om}}
\end{figure}

\begin{figure*}
\begin{center}
\end{center}
\caption{Color-magnitude and color-color diagrams for all OM sources. 
Large dots with error bars correspond to the closest counterparts of the X-ray sources 
(see text).
\label{colmag}}
\end{figure*}

\subsection{Optical Monitor data}

During the X-ray observation, the field was also observed through 
two UV filters (UVW1 and UVM2) and the visible (V) filter of the 
Optical Monitor (OM) onboard {\it XMM-Newton} (Mason et al. \cite{mas}). 
The exposure times were 1~ks, 3~ks, and 1~ks for the UVW1, UVM2, 
and V filters, respectively. The coverage of the full OM field 
necessitated five consecutive exposures. The OM is a 30 cm 
optical and UV telescope with extremely sensitive detectors. It is working 
in a photon-counting mode, analogous to the way the X-ray detectors work.
The data obtained with the OM are thus quite different from the `standard'
ground-based data, and present particular features, like fixed
`modulo-8' noise and significant coincidence losses even for relatively 
faint objects (Mason et al. \cite{mas}). Further complications follow 
from the fact that the OM Point Spread Function (PSF) depends on the 
source brightness\footnote{ Note that the algorithm used to derive the 
photometry can only handle a single PSF. However, the errors due to
this approximation are certainly smaller than those due to the 
crowdeness of the field.}.

We used SAS tasks $omatt$ and $ommosaic$ to create astrometrically 
corrected composite images of the full field-of-view (FOV). An 
example of such a composite image is shown in Fig. \ref{om} for 
the UVM2 filter. Note that it does not completely cover the entire EPIC FOV. 
A first estimate of the sky coordinates was obtained using the nominal 
pointing position of the telescope, and this may result in a systematic 
shift of our sky coordinates relative to the ideal ones by up to 4\arcsec.

Because of the particular features of the OM and of the 
extremely crowded nature of the field, the SAS tasks were unable to 
fully process the data and generate accurate photometry. For this reason, 
the photometric reduction was done with the {\rm daophot} program 
in the PSF fitting mode. The sources were first detected 
using the $sextractor$ task. The resulting list of stars was cleaned of 
spurious identifications, and some missing sources were added: a total 
of 6250 sources were then detected in this field. Next, photometry 
was derived for these stars using PSF-fitting. The zero point was 
determined by selecting relatively isolated stars and performing aperture 
photometry with an aperture radius of 3\arcsec. After the PSF 
magnitudes were adjusted to this zero point, they were scaled to 
the default aperture (6\arcsec) using information about the OM PSF 
from the SAS current calibration file OM\_PSF1DRB\_0006.CCF. Next, 
the empirical coincidence loss correction was applied and finally,
instrumental V magnitudes were converted to the standard V 
magnitudes. These transformations are based on the 
UVW1$-$V$^{{\rm OM}}$ color and were derived by the OM calibration team 
from simulations of the color transformations, using OM  instrumental 
response curves and several spectral libraries: V = V$^{{\rm OM}} - 0.0165 - 
0.0059*($UVW1$-$V$^{{\rm OM}}) - 0.0009*($UVW1$-$V$^{{\rm OM}})^2$.
In V, UVW1 and UVM2, the zero points of the OM are set to 17.9633, 
17.2965 and 15.8098, respectively. The corresponding simulated 
magnitudes of Vega are equal to U(Vega)=0.025~mag and V(Vega)=0.03~mag
(for more details about the OM calibration, see Antokhin et al., in 
preparation). 

To get a reliable astrometric solution for the OM sources, we used 
the GSC 2.2 catalog. We first made a tangential projection of all GSC 
sources to create a linear scale approximately resembling the scale 
of the OM. Then, we manually cross-identified a few tens of the 
brighter stars from GSC and OM plots, and used the IRAF's tasks 
$geomap$ to compute a non-linear polynomial transformation from the 
OM coordinate system to the GSC tangential one. Using this first 
approximation to cross-identify more sources, we then compute a more 
precise transformation (rms=0.22\arcsec), which gives reliable results 
for the GSC 2.2 catalog, the 2MASS All Sky Catalog of Point Sources 
and the USNO B1.0 catalog in the whole 
field. Note however that the transformation is less precise at the edges,  
so that we used an adaptive radius to get the actual GSC, 2MASS or USNO 
counterpart to the OM sources.

The OM astrometry and photometry of the $>$6000 sources  detected in 
\n\ are presented in an electronic table that can   be retrieved from 
the Centre de Donn\'ees Astronomiques de Strasbourg (CDS). In this table, 
the first column gives the OM number of the source; the second and 
third columns show the OM xy coordinates; the next seven columns 
provide the J2000 right ascension and declination of the sources; columns 
11 to 16 present the OM magnitudes in each filter, together with their 
error; column 17 is `y' if the source has a 2MASS 
entry\footnote{2MASS designations are not included since they are 
simply the coordinates of the sources.}; columns 18 to 23 yield the 
name and distance to the counterpart from the GSC, USNO B1.0, and PGMW 
catalogs, respectively. 

Color-magnitude and color-color diagrams of all OM sources are shown in 
Fig. \ref{colmag}. These diagrams show a well marked main-sequence in 
both cases. Our V photometry is one magnitude deeper than PGMW and 
appears complete down to V=20~mag, which corresponds to an early A-type 
star on the main sequence, whereas the brightest stars rather have an 
O3V type. The color-magnitude diagrams of LH9, LH10, and the whole 
field are very similar, as was shown in classical UBV photometry by PGMW. 

\begin{figure}[htb]
\begin{center}
\end{center}
\caption{X-ray contours overlaid on a DSS image of \n. 
The combined X-ray data of both EPIC MOS have been binned to obtain pixels of
2.5\arcsec, and then smoothed by convolution with a gaussian ($\sigma=3$~px).
The contours are at levels 0.35, 0.5, 0.8, and 2 cts pixel$^{-2}$.
N11B (containing LH10) is the bright nebula just northeast of the 
center, while N11C (LH13) and N11D (LH14) 
are the two easternmost nebulae of the field.
\label{dsstot}}
\end{figure}

\section{X-ray sources in the field}

\subsection{List of sources}

The X-ray data of \n\ reveal several point sources and extended 
emissions, as can be seen in Fig. \ref{color}.  
The latter category will be extensively discussed in Sect. 4, 
and we will focus here only on the point sources.

To search for discrete X-ray sources, we have applied the detection 
metatask {\it edetect\_chain} simultaneously to the data from the
three EPIC cameras. 
We used three energy bands: $0.4-1.0$ keV, $1.0-2.0$ keV and $2.0-10.0$ keV. 
We eliminated the false detections mainly due to structures in the 
diffuse emissions by rejecting the extended sources, the sources with 
a logarithmic likelihood $<5$ in one detector and/or a combined logarithmic
likelihood for the three detectors $<30$. This procedure left 37 point sources in 
the field. Their positions are shown in Fig. \ref{totfield} and their properties are presented in Table \ref{crate}, in order of increasing right ascension. 
Their background-subtracted and vignetting-corrected count rates in 
each of the EPIC cameras are also indicated, together with Hardness Ratios 
(HRs) calculated from EPIC pn data. These HRs are defined as 
$HR1=(M-S)/(M+S)$ and $HR2=(H-M)/(H+M)$, where $S$, $M$, and $H$ 
correspond to the count rates in the $0.4-1.0$, $1.0-2.0$ and $2.0-10.0$ 
keV bands, respectively. Note that
the SAS detection tasks do not apply a correction for out-of-time events or 
dead times. For the EPIC pn data used in extended full frame mode, 
this implies an overestimate of the exposure time (and thus an 
underestimation of the count rates) by 2.3\%. Since this amount is well 
within the errors, we choose not to correct the pn count-rates 
provided by {\it edetect\_chain}. 
Finally, a few sources associated with the LH10 cluster were excluded by the
selection criteria defined above. They are thus not included in Table 
\ref{crate} but will be extensively discussed in Sect. 4.4.

\begin{sidewaystable*}
\begin{center}
\caption{Coordinates of the detected X-ray point sources with their 
individual count rates for each of the three detectors, in the 
$0.4-10$~keV range. A missing value means the source is in a gap or out of
the field-of-view for this particular detector. The Hardness Ratios 
(HRs) were evaluated on the EPIC pn data.
The right-hand side of the table shows the results of the correlation 
with the GSC 2.2 catalog, the 2MASS All-Sky Catalog of Point Sources, 
and the USNO B1.0 catalog. In these last columns, Nr represents
the number of counterparts in the correlation region, while d corresponds
to the distance between the X-ray source and its counterpart, if unique.
 \label{crate}}
\tiny
\medskip
\begin{tabular}{l l | c c c c c | c c c c | c c c c c | c c c c c c c } 
N& Name& CR MOS1& CR MOS2& CR pn & HR1 & HR2 & \multicolumn{4}{c|}{GSC 2.2} & \multicolumn{5}{c}{2MASS}& \multicolumn{7}{|c}{USNO B1.0}\\ 
& & 10$^{-3}$ cts s$^{-1}$ & 10$^{-3}$ cts s$^{-1}$ & 10$^{-3}$ cts s$^{-1}$ & & & Nr & d(") & R & B & Nr & d(") & J & H & K & Nr & d(") & B1& R1 & B2 & R2 & I\\   
\hline
1& XMMU J045349.8$-$661928& & & 9.9$\pm$1.6 & 0.44$\pm$0.14 & -0.07$\pm$0.19 &     1 &2.9&  & 18.6  & 0& & &  & & 1&3.7 & & & 18.4  &18.9&\\
2& XMMU J045404.6$-$661657& & & 9.1$\pm$1.5 & 0.26$\pm$0.14 & -0.41$\pm$0.25 &     1 &3.4 & 17.9 & & 0& & &  & &1 &2.1   && 17.2  &18.1 & 16.9 & 17.1\\
3& XMMU J045423.4$-$663349& 2.4$\pm$0.7 & 1.1$\pm$0.5 & 3.9$\pm$1.0 & 0.30$\pm$0.25 & -0.22$\pm$0.34 &     1  & 3.4 && 13.8 & 0& &  & & & 1&3.8  && 18.5 & 18.1 & 18.6&\\
4& XMMU J045427.5$-$662522$^a$ & 2.2$\pm$0.6 & 1.9$\pm$0.5 & & & &0 & & & & 0 & & & & &2& & & & & & \\
5& XMMU J045429.4$-$661510& & & 15.6$\pm$2.2 & 0.32$\pm$0.14 & -0.05$\pm$0.17 &0 & & & & 0 & & & & &0 & & & & & & \\
6& XMMU J045434.2$-$661537 & 2.9$\pm$0.9 & & 4.7$\pm$1.2 & -0.12$\pm$0.29 & 0.35$\pm$0.28 &0 & & & & 0 & & & & &1 &2.9 && 18.6 & 20.1  &18.6  &\\
7& XMMU J045446.4$-$662004 & 5.9$\pm$0.7 & 7.4$\pm$0.8 & 16.8$\pm$1.4 & 0.51$\pm$0.09 & 0.18$\pm$0.08 &     1  & 1.0 & 18.2 &17.9 & 0& & & & &4&  & & & & & \\
8& XMMU J045448.9$-$663120 & 1.7$\pm$0.5 & 2.3$\pm$0.5 & 6.5$\pm$0.9 & 0.27$\pm$0.18 & 0.14$\pm$0.14 &0 & & & & 0 & & & & &0& & & & & & \\
9& XMMU J045454.6$-$663733 & 2.4$\pm$0.7 & 1.9$\pm$0.7 & 7.5$\pm$1.3 & 0.19$\pm$0.28 & 0.42$\pm$0.17 &0 & & & & 0 & & & & &1 &4.1  && 18.4 & 20.5 & 18.2 & 17.7\\
10& XMMU J045501.8$-$663123 & 0.8$\pm$0.3 & 1.1$\pm$0.3 & 5.3$\pm$0.7 & -0.27$\pm$0.12 & -0.86$\pm$0.24$^g$ &     1  & 1.7 & 14.4 &    &1& 1.8  & 13.9  & 13.5   & 13.4 
&2& & & & & & \\
11& XMMU J045502.0$-$663147 & 1.7$\pm$0.4 & 1.4$\pm$0.3 & 3.1$\pm$0.7 & 0.64$\pm$0.29$^e$ &  0.00$\pm$0.21 &     1 &  0.3 &17.5&  & 0& & & & &0& & & & & & \\
12& XMMU J045502.2$-$663349 & 0.9$\pm$0.4 & 1.1$\pm$0.4 & 3.1$\pm$0.8 & 0.22$\pm$0.53$^e$ & 0.57$\pm$0.21 &     1 &  3.0 &&18.0  & 0& & & & &1 &2.8&&&17.5 & 18.1  &  \\
13& XMMU J045509.0$-$663018 & & 1.9$\pm$0.4 & 3.9$\pm$0.6 & 1.00$\pm$0.09$^e$ & 0.43$\pm$0.13 &    1 &  0.9&17.3 & & 1&1.4 &16.0  &15.2 & 15.3 &1 &1.7&&16.7&15.8 &16.7&\\
14& XMMU J045513.4$-$663225 & 1.9$\pm$0.4 & 1.4$\pm$0.3 & 3.6$\pm$0.7 & 0.52$\pm$0.38$^e$ & 0.43$\pm$0.16 &0 & & & & 0 & & & & &1 &4.2&& 17.4 &15.7&17.8&\\
15& XMMU J045515.1$-$663039 & 2.2$\pm$0.5 & 3.4$\pm$0.9 & 4.6$\pm$0.7 & 0.77$\pm$0.21$^e$ & 0.08$\pm$0.14 &0 & & & & 0 & & & & &0& & & & & & \\
16& XMMU J045517.8$-$661736 & 3.5$\pm$0.6 & 4.6$\pm$0.7 & 8.6$\pm$1.0 & 0.45$\pm$0.14 & 0.01$\pm$0.13 &0 & & & & 0 & & & & &0& & & & & & \\
17& XMMU J045524.6$-$662224 & 1.7$\pm$0.3 & 2.3$\pm$0.3 & 4.7$\pm$0.7 & 0.77$\pm$0.16 & -0.23$\pm$0.14 &0 & & & & 0 & & & & &0& & & & & & \\
18& XMMU J045533.8$-$663015 & 0.7$\pm$0.2 & 0.6$\pm$0.2 & 1.4$\pm$0.4 & & 0.89$\pm$0.21$^{e,f}$ &0 & & & & 0 & & & & &0& & & & & & \\
19& XMMU J045534.9$-$661525$^b$ & 3.2$\pm$0.5 & 4.8$\pm$0.7 & 12.6$\pm$1.2 & -0.72$\pm$0.06 & -1.00$\pm$0.28$^g$ &     1 &  1.0&& 11.5  &            1&1.2& 9.2  & 8.7 & 8.7 &1 &1.0&11.4 & 10.1  &10.9  &10.1  & 9.8\\
20& XMMU J045536.4$-$662526 & 0.6$\pm$0.2 & 1.0$\pm$0.2 & 2.6$\pm$0.4 & 1.00$\pm$0.12$^e$ & 0.21$\pm$0.16 &0 & & & & 0 & & & & &1 &4.4&&& 20.2 && 17.4\\
21& XMMU J045538.8$-$662549 & 1.1$\pm$0.2 & 1.4$\pm$0.2 & 2.3$\pm$0.6 & 1.00$\pm$0.15$^e$ & -0.05$\pm$0.26 &0 & & & & 0 & & & & &0& & & & & & \\
22& XMMU J045539.6$-$662959 & 5.6$\pm$0.5 & 5.6$\pm$0.5 & 17.4$\pm$1.1 & 0.14$\pm$0.08 & -0.02$\pm$0.07 &     1  & 2.9& 16.4 & & 0& & & & &2& & & & & & \\
23& XMMU J045554.7$-$662715 & 1.8$\pm$0.3 & 1.9$\pm$0.3 & 5.6$\pm$0.6 & 0.48$\pm$0.15 & -0.06$\pm$0.11 &0 & & & & 0 & & & & &0& & & & & & \\
24& XMMU J045555.0$-$661701 & 0.9$\pm$0.3 & 0.9$\pm$0.4 & 4.4$\pm$0.8 & -0.35$\pm$0.14 & -0.31$\pm$0.34 & 0& & & & 1& 0.8 & 15.1&  14.5 &14.2 &0& & & & & & \\
25& XMMU J045619.2$-$662631 & 0.9$\pm$0.2 & 0.9$\pm$0.3 & 2.0$\pm$0.5 & 0.46$\pm$0.39$^e$ & 0.04$\pm$0.22 &0 & & & & 0 & & & & &0& & & & & & \\
26& XMMU J045624.3$-$662205 & 2.8$\pm$0.4 & 3.0$\pm$0.4 & 8.0$\pm$0.8 & 0.06$\pm$0.12 & -0.17$\pm$0.10 &0 & & & & 0 & & & & &0& & & & & & \\
27& XMMU J045629.7$-$662701$^c$ & 0.5$\pm$0.2 & 0.7$\pm$0.3 & 2.0$\pm$0.5 & 0.49$\pm$0.49$^e$ & 0.30$\pm$0.21 &0 & & & & 0 & & & & &2& & & & & & \\
28& XMMU J045632.2$-$661908 & 0.8$\pm$0.3 & 0.4$\pm$0.3 & 2.4$\pm$0.6 & 1.00$\pm$0.31$^e$ & 0.05$\pm$0.24 &0 & & & & 0 & & & & &0& & & & & & \\
29& XMMU J045637.1$-$661828 & 0.7$\pm$0.3 & 1.4$\pm$0.4 & 4.5$\pm$0.7 & 0.25$\pm$0.25 & 0.22$\pm$0.16 &0 & & & & 0 & & & & &1 &2.9&&16.9  &15.0 &&17.2\\
30& XMMU J045706.7$-$662705 & 0.8$\pm$0.4 & 1.1$\pm$0.3 & 3.4$\pm$0.7 & 0.41$\pm$0.25 & -0.16$\pm$0.21 &     1 &  3.5&16.3& & 0& & & & &2& & & & & & \\
31& XMMU J045727.9$-$662234 & 1.3$\pm$0.4 & 1.5$\pm$0.4 & 4.8$\pm$0.9 & 0.67$\pm$0.27$^e$ & 0.49$\pm$0.13 &0 & & & & 0 & & & & &0& & & & & & \\
32& XMMU J045728.1$-$663302 & 2.8$\pm$0.5 & & 8.1$\pm$1.2 & 1.00$\pm$0.41$^e$ & 0.57$\pm$0.11 &     1 &  2.5&  16.4& & 0& & & & &0& & & & & & \\
33& XMMU J045729.2$-$662256 & 1.1$\pm$0.4 & 1.5$\pm$0.4 & 2.7$\pm$0.7 & 1.00$\pm$0.25$^e$ & 0.41$\pm$0.23 &0 & & & & 0 & & & & & 0& & & & & & \\
34& XMMU J045732.2$-$662856 & 1.5$\pm$0.4 & 0.7$\pm$0.3 & 3.9$\pm$0.9 & 0.27$\pm$0.31 & 0.24$\pm$0.22 &0 & & & & 0 & & & & & 0& & & & & & \\
35& XMMU J045744.1$-$662741$^d$ & 3.9$\pm$0.6 & 2.5$\pm$0.5 & 10.4$\pm$1.2 & -0.15$\pm$0.11 & -1.00$\pm$0.14$^g$ &     1  & 0.9 &   &11.8 &            1& 1.2 & 11.9& 12.0& 11.9 &1 &0.9& 12.1  &13.0  &12.7 & 13.1 & 13.3\\
36& XMMU J045749.2$-$662726 & 1.6$\pm$0.4 & 0.5$\pm$0.3 & 3.1$\pm$1.0 & 0.00$\pm$0.39 & 0.28$\pm$0.35 &0 & & & & 0 & & & & &0& & & & & & \\
37& XMMU J045749.2$-$662210 & 1.4$\pm$0.4 & & 5.1$\pm$1.0 & 0.56$\pm$0.26 & 0.28$\pm$0.17 &0 & & & & 0 & & & & &0& & & & & & \\
\end{tabular}
\end{center}
\tiny
$^a$This source corresponds to BSDL 188 (see Fig. \ref{n11l} and Sect. 3.5). 
An infrared source, IRAS 04543$-$6629, is situated at 4\farcs1 from the X-ray
source, and may be coincident with it.\\
$^b$This source corresponds to HD 268670 (B=11.3,V=10.49, spectral type K0).\\
$^c$PGMW 2088 (B=18.6, V=18.8) lies at 1\farcs9 from this source.  \\    
$^d$This source corresponds to the tight cluster HD 268743 (Sk$-$66$^{\circ}$41, B=11.6, V=11.7, combined spectral type O3V((f*))+OB) \\
$^e$For this source, the count rate in the soft band is, within the errors, close to zero.\\
$^f$For this source, the count rate in the medium band is, within the errors, close to zero.\\
$^g$For this source, the count rate in the hard band is, within the errors, close to zero.\\
\end{sidewaystable*}
\normalsize

\subsection{Identification}

We have compared our source list to the {\it ROSAT} detections in the same 
field. Six catalogs are available: Haberl \& Pietsch (\cite{hab}, [HP99]), 
Sasaki et al. (\cite{sas}, [SHP2000]), the WGA catalog (White et 
al. \cite{whi}, 1WGA) and three catalogs from the {\it ROSAT} teams 
({\it ROSAT} consortium \cite{rxh}a, 1RXH; {\it ROSAT} consortium 
\cite{rxp}b, 2RXP; and Voges et al. \cite{vog99} \&  \cite{vog00}, 1RXS). 
The source detection algorithms used in the last four 
catalogs were most probably confused  by the presence of diffuse 
emissions and generated a lot of false detections. They should thus 
be taken more cautiously than the first two and are only given here 
for information. The cross-correlations between \x\ and {\it ROSAT} 
source names are presented in Table \ref{oldx}. For completeness, the 
positional error on the {\it ROSAT} detections and their separation 
from the \x\ sources are also listed in the last two columns of 
the Table. Generally, the sources' positions agree within the errors. 
There seems to be no bright transient source in the field of \n.

To search for optical counterparts, we have cross-correlated our source 
list with the Simbad catalog, the USNO B1.0 catalog, the 2MASS All-Sky 
Catalog of Point Source and the GSC 2.2 catalog. We have defined a candidate 
counterpart as an object lying less than 5\arcsec\ (the FWHM of the \x\ 
PSF) from the X-ray source. 21 X-ray sources may possess such visible 
counterparts, and we give the latter in the right-hand side of Table 
\ref{crate}, together with the separation between the X-ray sources
and these objects, 
and some basic informations about them. Fig. \ref{dsstot}
presents X-ray contours overlaid on a DSS image of the field. Note that 
the LH14 cluster is not included in the {\it XMM} FOV. Our 5\arcsec\ 
criterion failed to associate XMMU J045744.1$-$662741 (source \#35) 
with Sk $-66^{\circ}$41, 
due to the low precision of the coordinates of this object in the  
catalog of Sanduleak (\cite{san})\footnote{Note that if we use the coordinates 
found in the USNO, GSC, and 2MASS catalogs, the X-ray source is less than 
1\farcs2 away from the cluster core.}. However, as there is little doubt about
this identification (see Sect. 5), it was added to Table \ref{crate}.
We also note that XMMU J045538.8$-$662549 (source \#21) is quite close to a bright Cepheid (see below). 

Counterparts were also identified from the OM data. Using our
astrometric solution for the OM sources (see Sect. 2.2), 
we correlated the OM source list with the X-ray one. Table \ref{ommag} 
gives the photometry of the OM counterpart(s) found within 5\arcsec\ 
of our X-ray sources. 
The optical counterparts of the X-ray sources can be compared
  with field sources in the color-magnitude and color-color
  diagrams presented in Fig. \ref{colmag}. 
Even if the OM  filter combination used in our observation is not optimal 
for classification purposes (see 
Royer et al. \cite{roy}), the colors apparently favor a stellar 
origin for most of these counterparts. The majority lies within the main 
sequence of the clusters, but the counterpart of XMMU J045501.8$-$663123
(source \#10) is a rather bright red star lying clearly outside the main 
sequence. Only a very small number of these OM 
counterparts may possibly be quasars at low redshift, but quasars at 
high redshift are highly improbable.

\begin{table}[htb]
\caption{X-ray sources previously detected in the field. 
The number in the first column refers to our internal numbering 
scheme (see Table \ref{crate}).
Identifications with the 2RXP and 1WGA catalogs
have only an indicative value. When there were two entries 
with the same name in the 1WGA catalog, we kept only the 
source detected with the smallest off-axis angle, as
recommended by the WGA team.
The second part of the Table deals with the diffuse extended
emissions. Several spurious X-ray sources were
detected in the 2RXP and 1WGA catalogs at their position,
and we will not list them here. 
\label{oldx}}
\begin{center}
\begin{tabular}{l l c c} 
\hline
N & {\it ROSAT} name & Pos. error& Sep.\\
& & (\arcsec) & (\arcsec) \\
\hline
  5& 2RXP J045431.1$-$661530 & 15 &22.2\\
  6& 2RXP J045431.1$-$661530 & 15 &20.8\\
  9& $\left\{ \begin{array}{l} {\rm [HP99]\,\,394} \\ {\rm 2RXP \,\,J045455.2-663731} \end{array} \right. $  & $\left. \begin{array}{c} 18.1 \\ 8 \end{array} \right. $ & $\left. \begin{array}{c} 2.7 \\ 4.0 \end{array} \right. $\\
 10& 2RXP J045502.5$-$663126 & 8 &  4.2\\
 15& 2RXP J045516.1$-$663041 & 4 &  5.9\\
 16& 2RXP J045518.4$-$661738 & 5 & 3.8\\
 17& 2RXP J045523.9$-$662232 & 11 &  8.3\\
 19& $\left\{ \begin{array}{l} {\rm 2RXP \,\,J045535.4-661522} \\ {\rm 1WGA \,\,J0455.6-6615} \end{array} \right. $  & $\left. \begin{array}{c} 7 \\ 13 \end{array} \right. $ & $\left. \begin{array}{c} 4.8 \\ 13.5 \end{array} \right. $\\
 21& 1WGA J0455.6$-$6626 & 50. & 35.8\\
 22& $\left\{ \begin{array}{l} {\rm [SHP2000]\,\,LMC 14} \\ {\rm 2RXP \,\,J045540.5-662958} \\ {\rm 1WGA \,\,J0455.6-6629} \end{array} \right. $  & $\left. \begin{array}{c} 7.7 \\ 4 \\ 13 \end{array} \right. $ & $\left. \begin{array}{c} 1.4 \\ 5.3 \\ 6.1 \end{array} \right. $\\
 24& $\left\{ \begin{array}{l} {\rm 2RXP \,\,J045553.6-661709} \\ {\rm 2RXP \,\,J045555.3-661715} \end{array} \right. $  & $\left. \begin{array}{c} 11 \\ 12 \end{array} \right. $ & $\left. \begin{array}{c} 11.0 \\ 13.5 \end{array} \right. $\\
 26& 1WGA J0456.3$-$6622 & 13 &  7.0\\
 35& $\left\{ \begin{array}{l} {\rm 2RXP \,\,J045744.0-662739} \\ {\rm 2RXP \,\,J045744.9-662744} \\ {\rm 1WGA \,\,J0457.7-6627} \end{array} \right. $  & $\left. \begin{array}{c} 4 \\ 6 \\ 13 \end{array} \right. $ & $\left. \begin{array}{c} 2.3 \\ 5.6 \\ 6.3 \end{array} \right. $\\
\hline
N11L & $\left\{ \begin{array}{l} {\rm [SHP2000] LMC 13} \\ {\rm [HP99] 329} \end{array} \right. $ & $\left. \begin{array}{c} 8.3 \\ 4.8 \end{array} \right. $ & \\
LH9 & [HP99] 345 & 12.2& \\
\hline
\end{tabular}
\end{center}
\end{table}

\subsection{Variability}

We have analysed the lightcurves of the 6 brightest point sources (i.e. with at 
least 50 cts in EPIC MOS and 100 cts in EPIC pn in the $0.4-10.0$ keV range). 
The count rates in each bin were background-subtracted using annuli around
the sources or close-by circles for the background estimation regions. 
The effective bin lengths were calculated 
by taking into account the `good time intervals' defined in Sect. 2. The 
resulting lightcurves were analysed using Kolmogorov-Smirnov and $\chi^2$ 
tests. We also ran a modified probability of variability test 
(Sana et al. \cite{sana}) 
on the event lists of these sources. In the \n\ field, no source was found 
significantly variable over the time of our observation.

\begin{table}[htb]
\caption{Photometry of the OM counterparts lying within 5\arcsec\ 
from X-ray sources.
The number in the first column refers to our internal numbering 
scheme (see Table \ref{crate}).
\label{ommag}}
\begin{center}
\begin{tabular}{l c c c c} 
\hline
N & V$\pm\sigma_{{\rm V}}$ & UVW1$\pm\sigma_{{\rm UVW1}}$ &UVM2$\pm\sigma_{{\rm UVM2}}$ & Sep.\\
& (mag)& (mag)&(mag)&(\arcsec) \\
\hline
10 &$15.57\pm0.04$   &$17.56\pm0.07$  &$19.24\pm0.26$ & 1.7  \\ 
11 &$19.34\pm0.18$   &$19.20\pm0.15$  &$18.78\pm0.25$ & 1.8   \\
13 &$17.72\pm0.06$   &$19.43\pm0.15$  &$20.02\pm0.34$ & 1.0  \\
14 &$19.76\pm0.19$   &$19.30\pm0.16$  &$19.22\pm0.21$ & 4.4\\
17 & $\left\{ \begin{array}{c} 19.84\pm0.17 \\ 20.00\pm0.16 \end{array} \right. $  & $\left. \begin{array}{c} 19.38\pm0.22 \\ 19.43\pm0.19  \end{array} \right. $ & $\left. \begin{array}{c} 18.86\pm0.17 \\ 18.56\pm0.18 \end{array} \right. $& $\left. \begin{array}{c} 2.8 \\  3.7 \end{array} \right. $ \\
20 &$17.25\pm0.06$   &$16.23\pm0.03$  &$16.04\pm0.03$ & 4.4 \\
22 &$19.27\pm0.13$   &$18.34\pm0.13$  &$19.00\pm0.15$ & 0.2   \\
27 & $\left\{ \begin{array}{c} 19.20\pm0.13 \\ 19.44\pm0.20 \end{array} \right. $  & $\left. \begin{array}{c} 19.28\pm0.18 \\  19.23\pm0.19 \end{array} \right. $ & $\left. \begin{array}{c} 19.73\pm0.28 \\ 18.89\pm0.17 \end{array} \right. $& $\left. \begin{array}{c} 1.8 \\ 4.5  \end{array} \right. $ \\
28 &$18.98\pm0.13$   &$19.74\pm0.21$  &$19.90\pm0.32$ & 3.6   \\
30 &$16.54\pm0.05$   &$14.89\pm0.03$  &$14.60\pm0.02$ & 3.7  \\
\hline
\end{tabular}
\end{center}
\end{table}

\subsection{Spectral properties}

We have also extracted\footnote{To ensure a correct determination of the X-ray 
properties of the sources, all extractions (spectra, lightcurves) were made in 
detector coordinates.} the spectra of these brightest 
sources. We have generated response matrix files (rmf) and ancillary 
response files (arf)
using the SAS tasks {\it rmfgen} and {\it arfgen}. The spectra were 
 binned to reach a minimum of 10 cts per channel. 
Finally, we analysed the background-corrected spectra 
within XSPEC. Due to strong noise at very low and very high energies, we have
discarded bins below 0.4 keV or above 10 keV. We have fitted the spectra
with an absorbed $mekal$ model or an absorbed power-law. For each source,
we have fitted separately EPIC MOS1+2 and EPIC pn data, but as they
gave similar results -- within the errors -- we finally fitted all three
instruments at the same time and we list the parameters of the best-fit 
models in Table \ref{spec}.

\subsection{Individual sources}
\begin{table*}
\begin{center}
\caption{Parameters of the spectral fits for the brightest point sources. 
Stated errors  correspond to the 90 \% confidence level. The global metallicity
was set to $0.3\,Z_{\odot}$, as appropriate for the LMC, for all sources except 
XMMU J045534.9$-$661525 (source \# 19), for which we adopt solar abundances since it 
corresponds to a foreground star. Absorbed fluxes are given in the spectral 
range $0.4-10$~keV and in units 10$^{-14}$ergs cm$^{-2}$ s$^{-1}$. The results 
correspond to a simultaneous fit of all EPIC spectra available. A constrained 
fit for XMMU J045744.1$-$662741 is also listed below.
\label{spec}}\medskip 
\begin{tabular}{l c c c c | c c c c} 
\hline
N & \multicolumn{4}{c|}{Parameters of the $mekal$ model} & \multicolumn{4}{c}{Parameters of the absorbed power law}\\
 & $N($H) & $kT$ & $\chi^2_{\nu}$(dof) & $f_X^{abs}$  & $N($H) & $\Gamma$ & $\chi^2_{\nu}$(dof) & $f_X^{abs}$ \\
 & (10$^{22}$ cm$^{-2}$) & (keV) &  &  & (10$^{22}$ cm$^{-2}$)&   & &  \\
\hline
\vspace*{-0.3cm}&&&&&&&&\\
7 & 0.21$_{0.07}^{0.43}$ & 14.2$_{4.2}^{100.}$ & 1.01 (47) & 6.81 & 0.33$_{0.16}^{0.66}$ & 1.68$_{1.29}^{2.23}$  & 0.98 (47)& 6.55\\
\vspace*{-0.3cm}&&&&&&&&\\
16 & 0.36$_{0.07}^{0.76}$ & 6.40$_{2.68}^{100.}$ & 1.50 (29) & 4.33 & 0.45$_{0.20}^{1.03}$ & 1.80$_{1.47}^{2.69}$ & 1.49 (29)& 4.57\\
\vspace*{-0.3cm}&&&&&&&&\\
19 & 0.45$_{0.}^{0.66}$ & 0.24$_{0.18}^{0.31}$ & 1.08 (41) & 1.99 & 1.11$_{1.03}^{1.19}$ & 10.$_{8.40}^{10.}$ & 1.30 (41)& 1.95\\
\vspace*{-0.3cm}&&&&&&&&\\
22 & 0.05$_{0.}^{0.17}$ & 7.14$_{4.16}^{20.7}$ & 0.82 (66) & 5.93 & 0.10$_{0.02}^{0.26}$ & 1.68$_{1.44}^{2.02}$ & 0.84 (66)& 6.28\\
\vspace*{-0.3cm}&&&&&&&&\\
26 & 0.87$_{0.55}^{1.24}$ & 0.76$_{0.57}^{1.10}$ & 0.99 (58) & 1.19 & 0.37$_{0.14}^{0.64}$ & 3.25$_{2.69}^{4.44}$ & 1.01 (58)& 1.56\\
\vspace*{-0.3cm}&&&&&&&&\\
35 & 0.$_{0.}^{0.22}$ & 0.84$_{0.73}^{1.10}$ & 1.16 (19) & 1.85 & 0.14$_{0.04}^{0.36}$ & 3.47$_{2.79}^{4.18}$ & 0.91 (19)& 2.84\\
\vspace*{-0.3cm}&&&&&&&&\\
\hline
\vspace*{-0.3cm}&&&&&&&&\\
35$^1$& 0.06$_{0.06}^{0.41}$ & 0.94$_{0.58}^{1.09}$ & 1.29 (19) & 1.78 &  &  &  & \\
\vspace*{-0.3cm}&&&&&&&&\\
\hline
\end{tabular}
\end{center}
$^1$ $N($H) constrained to the interval $0.06-0.6\times10^{22}$ cm$^{-2}$ (see Sect. 4.1).\\
\end{table*}

\begin{table*}
\begin{center}
\caption{Parameters of the spectral fits for the extended sources. 
Stated errors  correspond to the 90 \% confidence level. The global
metallicity was set to $0.3\,Z_{\odot}$ for all sources. Fluxes are 
given in the spectral range $0.4-10$~keV, and in units 10$^{-14}$ergs 
cm$^{-2}$ s$^{-1}$. Unabsorbed fluxes are corrected for the fitted $N($H). 
The results correspond to a simultaneous fit of all EPIC spectra, except 
for the NE source, for which only a pn spectrum was available. Constrained 
fits are also listed below when the fitted $N($H) was outside the `allowed' 
range (see Sect. 4.1). Note that solutions with $\chi^2>2$ were not included.
\label{specdif}}\medskip 
\begin{tabular}{l c c c c c | c c c c c } 
\hline
Source & \multicolumn{5}{c|}{Parameters of the $mekal$ model} & \multicolumn{5}{c}{Parameters of the absorbed power law}\\
 & $N($H) & $kT$ & $\chi^2_{\nu}$(dof) & $f_X^{abs}$ & $f_X^{unabs}$ & $N($H) & $\Gamma$ & $\chi^2_{\nu}$(dof) & $f_X^{abs}$ & $f_X^{unabs}$ \\
 & (10$^{22}$ cm$^{-2}$) & (keV) &  &  & &(10$^{22}$ cm$^{-2}$)&   & & &\\
\hline
\vspace*{-0.3cm}&&&&&&&&&&\\
N11L& 0.43$_{0.40}^{0.46}$ & 0.15$_{0.15}^{0.16}$ & 1.63 (162) & 23.1 & 459. &  & & & & \\
\vspace*{-0.3cm}&&&&&&&&&&\\
plume N of N11L& 0.68$_{0.64}^{0.73}$ & 0.15$_{0.14}^{0.18}$ & 1.29 (175) & 11.0 & 755. & 0.72$_{0.57}^{0.92}$ & 7.38$_{6.37}^{8.76}$ & 1.79 (175) & 11.2 & 1500.\\
\vspace*{-0.3cm}&&&&&&&&&&\\
LH9& 0.41$_{0.37}^{0.45}$ & 0.18$_{0.17}^{0.18}$ & 1.54 (419) & 45.9 & 663. && &  & \\
\vspace*{-0.3cm}&&&&&&&&&&\\
LH10& 0.05$_{0.}^{0.17}$ & 0.59$_{0.52}^{0.66}$ & 1.39 (35) & 4.56 & 5.56 &   &  &  & \\
\vspace*{-0.3cm}&&&&&&&&&&\\
NE& 0.36$_{0.10}^{0.44}$ & 0.22$_{0.19}^{0.29}$ & 1.05 (45) & 4.56 & 34.0 & &  &  &  & \\
\vspace*{-0.3cm}&&&&&&&&&&\\
NW& 0.26$_{0.20}^{0.35}$ & 0.23$_{0.20}^{0.25}$ & 1.36 (154) & 9.29 & 40.8 & & &  &  & \\
\vspace*{-0.3cm}&&&&&&&&&&\\
\hline
\vspace*{-0.3cm}&&&&&&&&&&\\
LH9$^1$& 0.20$_{0.18}^{0.20}$ & 0.23$_{0.22}^{0.23}$ & 1.74 (419) & 45.0 & 100. &  &  &  &  & \\
\vspace*{-0.3cm}&&&&&&&&&&\\
LH10$^2$& 0.10$_{0.10}^{0.18}$ & 0.58$_{0.51}^{0.63}$ & 1.41 (35) & 4.49 & 6.51 &  &  &  &  & \\
\vspace*{-0.3cm}&&&&&&&&&&\\
\hline
\end{tabular}
\end{center}
$^1$ $N($H) constrained to the interval $0.06-0.2\times10^{22}$ cm$^{-2}$ (see Sect. 4.1).\\
$^2$ $N($H) constrained to the interval $0.1-0.8\times10^{22}$ cm$^{-2}$ (see Sect. 4.1).\\
\end{table*}

\begin{itemize}
\item XMMU J045427.5$-$662522 (source \# 4):\\
One of the most interesting discoveries in this field is the X-ray emission
associated with a small ellipsoidal object west of N11\,L, roughly 
28\arcsec$\times$\,24\arcsec in size. 
This extended object correponds to BSDL 188, also known 
in Simbad as MSX LMC 1241, which is classified as an emission nebula
plus association in Bica et al. (\cite{bic}).
But this object also correlates well with an IR source (IRAS 04543$-$6630,
or LEDA 89996, or QDOT B0454225$-$662949) classified as a nearby galaxy of 
radial velocity $\sim 7100-8500$ \kms (Lawrence et al. \cite{law}, LEDA 
database). At that position, two extended objects are also listed in the 
2MASS Extended Sources Catalog: 2MASX 04542820$-$6625275 (radius=29\arcsec)
and 2MASX 04542184$-$6625310 (radius=15\arcsec). Since Bica et al. (\cite{bic}) 
did only an imaging survey without any spectroscopic information, 
the BSDL 188 ``nebula" probably corresponds to this galaxy.  
Unfortunately, it lies outside the OM field-of-view and we thus cannot 
provide any information on the UV emission of this object. We note however 
that the X-ray source is not exactly centered on the extended 
object, but is slightly off-axis. We also remark that the X-ray source 
is not extended, as far as the resolution of 
{\it XMM-Newton} can tell. Deeper investigations of this peculiar 
source are needed to better understand the nature of the extended object 
and the underlying mechanisms of the X-ray emission.

\item XMMU J045446.4$-$662004 and XMMU J045517.8$-$661736 (i.e. our 
sources \# 7 and \# 16): \\
Their poor fit by a $mekal$ model, large hardness ratios, 
rather large $N($H$)$ (3$-$5$\times10^{21}$~cm$^{-2}$), and 
$\Gamma\sim 1.7-1.8$  suggest that these sources are
actually extragalactic background objects. Approximately one third of 
the point sources detected in this field possess large HRs, and probably 
do not belong to the LMC. 

\item XMMU J045534.9$-$661525 = HD 268670 (source \# 19):\\
This low-mass foreground K0 star is clearly detected in our
observation, with an X-ray flux of $\sim$2$\times$10$^{-14}$ ergs 
s$^{-1}$ cm$^{-2}$. No significant X-ray variability was found for 
this source. The X-ray spectrum of this object is well fitted by an 
optically thin thermal plasma model but not by a power law, in agreement 
with the idea of this being a coronal emission from a late-type star.

\item XMMU J045538.8$-$662549 (source \# 21):\\
This source is 21\arcsec\ away from a Cepheid star of type 
G2Iab. X-ray emission from Cepheids has never been detected 
(e.g. Evans et al. \cite{eva}), and it is not clear that 
the X-ray source is physically associated with the Cepheid. 
If it is the case, then a companion would most probably be 
responsible for the X-ray emission.

\end{itemize}

\begin{figure}[htb]
\begin{center}
\end{center}
\caption{ Regions of extraction for each diffuse source.  Circles of radius 30\arcsec\
centered on point sources XMMU J045619.2$-$662631 (source \#25), 
XMMU J045624.3$-$662205 (source \#26), and XMMU J045629.7$-$662701 
(source \#27) were excluded from the extractions. 
The northern extended emission was cut in two parts, since it lies 
on two different EPIC pn CCDs.
\label{dif}}
\end{figure}

\section{Diffuse emissions}

Soft diffuse emission is seen throughout the field (see Fig. \ref{color}).
The brightest and most compact diffuse X-ray emission region corresponds 
to the SNR N11L. Extended diffuse X-ray emission covers LH9, LH10 
and a region to the north of these clusters (see Figs. \ref{dsstot} and 
\ref{dif}). 
The emission associated with N11L, LH9 and LH10 will be discussed more 
extensively in Sects. 4.2, 4.3, and 4.4, respectively. The last region of 
diffuse emission, which will be called N11-North, is projected around 
XMMU J045624.3$-$662205 (i.e. source \# 26), but may not be physically 
connected to this source. This extended X-ray emission is confined by 
faint \ha\ filaments. It is  bisected by a north-south oriented ridge 
of low X-ray surface brightness coincident with an \ha\ filament; 
the low X-ray surface brightness is thus probably caused by a higher 
absorption column density provided by the filament. 
Meaburn et al. (\cite{mea}) and Mac Low et al. (\cite{mac}) studied 
the kinematics of N11 and revealed the existence of expanding motions 
in the \ha\ line at the position of the eastern part of N11-North 
(hereafter called `NE', see Fig. \ref{dif}), 
while they detected no line splitting further north. The X-ray 
emission may thus be the result of shocked expanding gas. Unfortunately, 
nothing is known on the kinematics of the western part of N11-North
(hereafter called `NW', see Fig. \ref{dif}).

To study these diffuse X-ray emission regions quantitatively, we have 
chosen the extraction regions shown in Fig. \ref{dif}. Nearby source-free 
regions were used as background for N11L and the extended X-ray emission 
to its north, but there is no emission-free region
near the other sources. Due to fluorescent features in the instrumental 
background that change with the position on the detector (see e.g.
Lumb \cite{lum}), we cannot use the background spectrum from a region
lying too far away from our diffuse sources for background subtraction. 
We thus decided to use blank sky event 
lists provided by Dr Andrew Read (Birmingham University). Since no blank 
sky files are available for the THICK filter, we have used the event 
lists associated with the MEDIUM filter. We note that there is only a slight
difference between the THIN and MEDIUM filters in these blank sky files 
for the regions considered. We also checked that using the MEDIUM blank sky 
for N11L and the extended X-ray emission 
to its north gives the same results as with a neighbouring 
background region. We thus have extracted blank sky spectra over the same 
detector regions as occupied by the diffuse emissions, then scaled 
these spectra to correctly subtract the fluorescence lines of the detector 
(this was only necessary once, for the MOS2 spectrum of LH9). 
 To generate rmf and arf matrices, we used the SAS tasks 
{\it rmfgen} and {\it arfgen} using a detector 
map of the region to account for the source extension. 
Note that as these extended emission regions cover generally more than one CCD, 
the response matrices are still to be taken with caution in SAS v5.4.1. 
The spectra were then binned to reach a minimum of 20 cts per channel. 
As we did for point sources, we discarded noisy bins: below 0.4 keV and above 
2 keV when the blank sky files were used or above 10 keV for N11L and the extended X-ray emission to its north. 
We list the parameters of the best-fit models in Table  \ref{specdif}. 
Note that they are consistent with {\it ROSAT} determinations (Mac Low et al.
\cite{mac}), but our new results have much smaller error bars.

\subsection{Hydrogen Columns}

\begin{figure}
\begin{center}
\end{center}
\caption{Close-up on N11L (DSS+EPIC MOS, as for Fig. 3). Contours
have been drawn at levels of 0.35, 0.5, 0.65, 0.8, 1, 1.5, 2, 2.5, and 3 
cts pixel$^{-2}$.  Note that a CCD gap ($\sim$10\arcsec in width) is 
present just east of the SNR. It 
crosses the field in the NNE-SSW direction, and can be located easily 
since it corresponds to a region of apparently lower X-ray emission.
The X-ray point sources present in the field are labeled.
\label{n11l}}
\end{figure}

\begin{figure*}
\begin{center}
\end{center}
\caption{The SNR N11L at different wavelengths: left, DSS red image; middle, image acquired through the OM UVW1 filter ($\lambda_0$=2298\,\AA\ and 
$\Delta(\lambda)$=439\,\AA) and right, image made in the OM UVM2 filter
($\lambda_0$=2905\,\AA\ and $\Delta(\lambda)$=620\,\AA). 
\label{omn11l}}
\end{figure*}

\begin{figure*}
\begin{center}
\includegraphics[width=8cm]{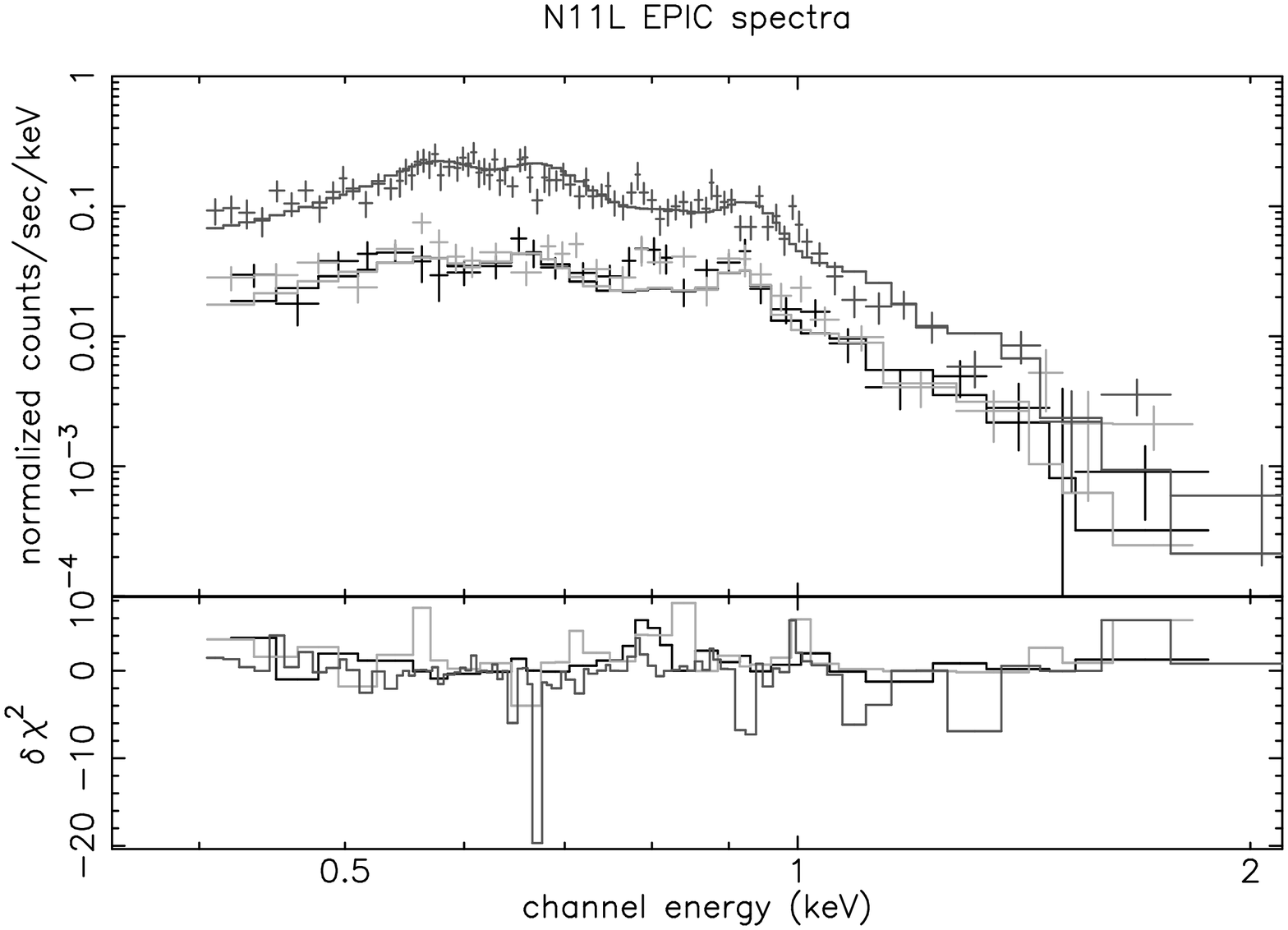}
\includegraphics[width=8cm]{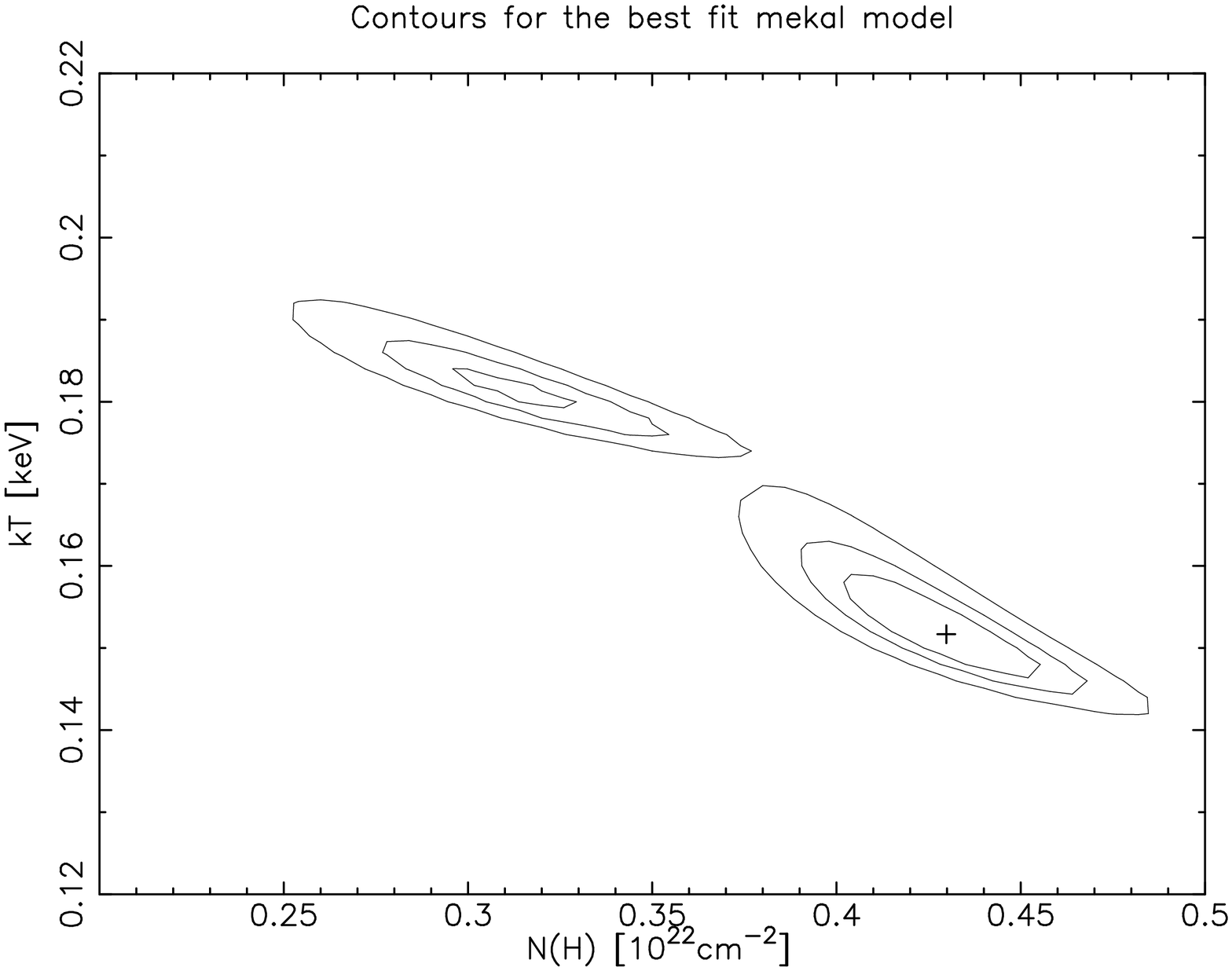}
\end{center}
\caption{Left: The X-ray spectrum of N11L, shown along with the best fit absorbed
{\em mekal} model, with $N($H$)=0.43\times 10^{22}$~cm$^{-2}$ and $kT=0.15$~keV 
(see Table \ref{specdif}). The upper data ( dark grey) come from the EPIC 
pn camera, the lower ones from EPIC MOS (black for MOS1 data and light 
grey for MOS2). Right: $\Delta\chi^2$ contours for the $mekal$ model: 
contours for 2D confidence levels of 68\%, 90\% and 99\% are drawn.
The cross indicates the model used in the fit of the left panel.
\label{specn11l}}
\end{figure*}

One parameter of the spectral models, the hydrogen column, can be estimated
using \hi/CO surveys, nebular emission, and stellar photometry. 
Rohlfs et al.\footnote{Note that Williams et al. (\cite{wil}) wrongly 
quoted the value from that paper, but this error did not affect their
 analysis since they finally used the \hi\ data from a high-resolution ATCA 
\hi\ survey.} (\cite{roh}) measured $N($\hi) of $1.8-2.4\times$10$^{21}$ 
cm$^{-2}$ in our {\it XMM} field-of-view, while Schwering \& Israel 
(\cite{sch}) quoted 5.6$\times$10$^{20}$ cm$^{-2}$ for the Galactic 
absorption column. The contribution of H$_2$ can also be estimated 
using the strength of the $^{12}$CO line (Israel et al. \cite{isr93}) 
and a conversion factor $N($H$_2)/I({{\rm ^{12}CO}})$ of $1.3-1.7 
\times$10$^{20}$ cm$^{-2}$ K$^{-1}$ km$^{-1}$~s. Israel et al. 
(\cite{isr93}) measured I$({{\rm ^{12}CO}})$ up to $18$ K km s$^{-1}$
for N11B and $14$ K km s$^{-1}$ for N11C. They also gave values in the 
range $7-21$ K km s$^{-1}$ in the $XMM$ field. The minimum absorption 
column towards \n\ is thus $6\times$10$^{20}$ cm$^{-2}$, while the 
maximum column could reach up to $1\times$10$^{22}$ cm$^{-2}$.
  
E(B$-$V) measurements (PGMW, Walborn et al. \cite{wal99}) and Balmer 
decrements (Caplan \& Deharveng \cite{cap86}, Heydari-Malayeri \& Testor 
\cite{hey85}, Heydari-Malayeri et al. \cite{hey87}) provide a second, 
independent estimate of $N($H). With a galactic extinction of $0-0.15$~mag 
towards the LMC (Oestreicher et al. \cite{oes}),
 gas-to-dust ratios of 5.8$\times$10$^{21}$ cm$^{-2}$ mag$^{-1}$
for the Galaxy (Bohlin et al. \cite{boh}) and 2.4 $\times$10$^{22}$ 
cm$^{-2}$ mag$^{-1}$ for the LMC (Fitzpatrick \cite{fit}), we get the 
following reasonable estimates of $N($H): $0.3-2\times$10$^{21}$ cm$^{-2}$ 
for LH9, $1-8\times$10$^{21}$ cm$^{-2}$ for LH10, $0.6-6\times$10$^{21}$ 
cm$^{-2}$ for LH13 or in general $0.3-8\times$10$^{21}$ cm$^{-2}$ for sources
actually belonging to N11. Combining all the constraints derived in this 
section, we fitted again the spectra when necessary (see Table 
\ref{specdif}).

\subsection{The SNR N11L}

N11L is an optically faint nebula of diameter $\sim$15~pc at the 
westernmost part of \n. It looks like a small filamentary bubble 
with a northeastern  extension. With an age of $10^4$ yr, this SNR 
expands with a mean velocity of $\sim$200\kms (Williams et al. 
\cite{wil}). The nonthermal radio emission 
of N11L is generally coincident with the \ha\ emission, but 
additional radio emission appears to the north of the main shell. 
{\it ROSAT} observations showed a rather faint X-ray source, compared 
to other LMC SNRs of similar size. It was cataloged as [SHP2000] LMC 
13 or [HP99] 329. In our \x\ observations (see Fig. \ref{n11l}), 
the X-ray emission is revealed in more details. It peaks more or less 
at the center of the shell, where the \ha\ (and  radio)
emission are lower. The X-ray emission is not completely confined 
to the \ha\ bubble: some X-ray emission is present just north
of that bubble, where radio emission is also enhanced. 
In addition, we discovered UV emission from N11L in the data 
from the OM telescope. The UV morphology of the SNR is very similar 
to the one in the \ha\ line (see Fig. \ref{omn11l}). 

The X-ray spectra of N11L are well fitted by a simple absorbed 
$mekal$ model. The parameters from the best fit are given in Table 
\ref{specdif}, but another solution, with a slightly larger $\chi^2$,
also exists, as can be seen in Fig. \ref{specn11l}. Note that
non-equilibrium models ($nei$ or $gnei$) do not give good fits to 
these data.

The {\it ROSAT} data suggested the presence of a large `plume' of 
diffuse emission extending at least 5\arcmin\ (or $\sim$70~pc at an LMC 
distance of 50~kpc) from the nebula. (Williams et al. \cite{wil}). 
This northern extension of diffuse X-ray emission is now clearly
detected in the EPIC data of \n.  Compared to N11L, the spectral
fit of this X-ray plume shows a similar temperature but a larger
extinction.  Note that its spectrum can also be fitted by a power-law, 
but that in this case the exponent $\Gamma$ is exceptionally large. 
The physical relationship between N11L and the X-ray
plume to its north is not clear.  N11L may contribute to some of
the hot gas in the plume region, but cannot be responsible for all of it, 
as the sound-crossing time ($>5\times10^5$yr for $c_S\sim$100\kms) 
of this X-ray plume is much larger than the age of the SNR itself
($\sim1.5\times10^5$yr, Williams et al. \cite{wil}).

\subsection{LH9}

The X-ray emission inside the central shell surrounding LH9 is thought to correspond to 
hot, shocked gas within a superbubble blown by the cluster stars. The 
kinematics of this region is complex, with several velocity components along 
our line-of-sight. However, Rosado et al. (\cite{ros}) found a general radial 
expansion velocity of $\sim$45\kms\ for this structure. The X-ray 
emission peaks near HD\,32228 (WC4+OB), where the most violent and complex 
gas motions are found. 

\begin{figure}
\begin{center}
\includegraphics[width=8cm]{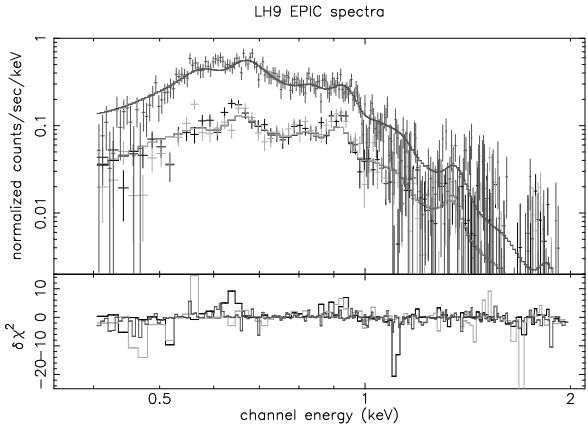}
\end{center}
\caption{The X-ray spectrum of the diffuse emission 
associated with LH9, shown along with the best fit absorbed
$mekal$ model. EPIC pn data are shown in dark grey, EPIC MOS1
in black and EPIC MOS2 in light grey.
\label{speclh9}}
\end{figure}

This brightest emission region, roughly 80\arcsec\ in radius around HD32228, 
fitted by the model derived for the whole region (see Table \ref{specdif} 
and Fig. \ref{speclh9}), presents an unabsorbed (i.e. corrected 
for the fitted $N($H)) luminosity of $\sim2-12\times$10$^{35}$ erg 
s$^{-1}$ in the $0.1-2.$ keV range ({\it ROSAT} range). In this region, 
15 (massive) stars were classified by Walborn et al. (\cite{wal99}) 
and 18 others by PGMW\footnote{[W99] 30, 89, 44 (the WC4 star), 62, 
21, 84, 50, 41, 51, 68, 61, 72, 36, 87, 
75, and PGMW 1052, 1110, 1125, 1160, 1194, 1200, 1239, 1288, 1292, 1303, 
1310, 1321, 1342, 1343, 1350, 1357, 1365, 1377}. The contribution 
of the WC4 star to the X-ray flux has been roughly estimated, since the 
X-ray emission 
from WC stars is still under debate. Ignace \& Oskinova (\cite{ign}) 
have found that (binary) WC stars have X-ray luminosities of 
$\sim$1.4$\times$10$^{32}$ erg s$^{-1}$. Pollock et al. (\cite{pol}) 
measured X-ray luminosities of $\sim$1$\times$10$^{32}$ erg s$^{-1}$ 
(WR38) or lower for Galactic (binary) WC4 stars, whereas following
Oskinova et al. (\cite{osk}), there has been up to now no conclusive 
detection of any single WC star. We adopt the highest value as an upper 
limit.  For the other stars, we use  $L_X^{unabs}(0.1-2.)/L_{{\rm BOL}}$ 
empirical relationships (Bergh\"ofer et al. \cite{ber}) to estimate their 
X-ray luminosities from the stellar photometric data published
by PGMW and Walborn et al. (\cite{wal99}) and the bolometric corrections
provided by Humphreys \& McElroy (\cite{hum}).  For the massive star population surrounding HD32228, we then expect a total unabsorbed X-ray luminosity of 
$\sim1.7\times10^{33}$ erg s$^{-1}$ in the $0.1-2.$ keV range.  
This is at least 100 times lower than observed.

For the whole cluster (i.e. in the same region as the one shown
in Fig. \ref{dif}), the discrepancy 
is even larger, with a predicted unabsorbed X-ray luminosity of 
$\sim$3$\times$10$^{33}$ erg s$^{-1}$ in the $0.1-2.$ keV energy range, for 
observations of $8-48\times$10$^{35}$ erg s$^{-1}$. In addition, 
we have to consider the role of the metallicity of the LMC. A lower 
metallicity $Z$ implies weaker winds, thus less X-ray emission will be
 produced, but the absorption by the wind is also lower in this case. 
Using {\it ROSAT} data of 
Galactic O stars, Kudritzki et al. (\cite{kud}) found that 
$L_X\propto (\dot M/v_{\infty})^{-0.38}\times L_{{\rm BOL}}^{1.34}$, i.e.
$L_X\propto Z^{\sim0.3}$ if we use the dependences of $\dot M$ and $v_{\infty}$ 
on $Z$ from Vink et al. (\cite{vin}) and the relationship between $\dot M$ and 
$L_{{\rm BOL}}$ from Howarth \& Prinja (\cite{how}). Our predictions
of the stellar X-ray luminosity may thus be slightly
overestimated, reinforcing the 
discrepancy with the observations. An additional source of X-ray emission 
is clearly needed. 

Shocked gas within the superbubble was suggested as the probable 
cause for this excess X-ray emission. Using the Weaver et al. (\cite{wea}) 
model, Dunne et al. (\cite{dun}) predicted an X-ray emission of 
$\sim$9$\times$10$^{34}$ erg s$^{-1}$ for the superbubble surrounding LH9. 
This reduces the discrepancy, but does not solve totally the conflict between 
observations and theoretical predictions. 

With an age of only 3.5 Myr 
(Walborn et al. \cite{wal99}), LH9 most probably still contains a lot
of Pre-Main Sequence (PMS) stars. With the development of recent X-ray 
observatories, it is now a common occurrence to serendipitously detect 
PMS objects in the field of OB clusters (e.g. Rauw et al. \cite{rau}, \cite{rau03}). 
Actually, the Classical and Weak Line T Tauri stars can present X-ray 
luminosities up to 10$^{31}$ erg s$^{-1}$. At the distance of the LMC, 
these objects would not be resolved as individual sources and could therefore 
mimic a diffuse emission. Using an IMF slope $\Gamma$ 
of $-1.6$ (PGMW), we find that $\sim$ 10$^4$ PMS stars of 
masses in the $0.2-2$ M$_{\odot}$ range might be lurking in the cluster, 
thus contributing to the X-ray luminosity for a maximum 10$^{35}$ erg~s$^{-1}$.

Another source of X-ray emission may come from a so-called 
`cluster wind'. For the Arches cluster, Cant\'o et al. (\cite{can}) have 
estimated that stars may constitute only 10\% of the X-ray emission, 
the hot intracluster gas being responsible for the rest. A similar
process might be at work in LH9, resulting in another increase of the
X-ray luminosity. Off-center supernova explosions hitting the swept-up shell 
and/or colliding wind binaries in the cluster may also provide additional 
sources of X-ray emission. More observations are needed to determine the 
importance of these two processes: the presence of these supernovae could be 
revealed by e.g.\ observations of high-velocity UV absorptions 
(Chu \& Mac Low \cite{cm}) whereas high-resolution spectroscopy of 
the binaries present in the cluster could enable to find the signatures of 
colliding winds interactions. 

Finally, we may note that soft X-ray emission is also present to the 
south of LH9 (see Fig. \ref{color}). This might indicate hot gas leaking 
out of the superbubble.

\subsection{LH 10}

With an age of $\sim$1Myr (Walborn et al. \cite{wal99}), LH10 is the 
youngest cluster of \n. Embedded in the bright nebula N11B, it contains 
several O3 stars, indicating that even the most massive stars have 
remained intact and that no SN explosions should have occurred in this 
cluster yet. In this pristine environment, the stars did not yet have 
the time to blow a superbubble surrounding the entire cluster. Nevertheless, 
Naz\'e et al. (\cite{naz}) showed that smaller wind-blown bubbles around 
individual stars or small groups of stars are present, e.g. near 
PGMW 3209 and around PGMW 3070.

\begin{figure}
\begin{center}
\end{center}
\caption{ HST WFPC2 image of N11B (see Naz\'e et al. \cite{naz}) superposed
onto X-ray contours (same contours as in Fig. \ref{n11l}). The most important 
stars are labelled with their PGMW number. Naz\'e et al. (\cite{naz}) found
expanding regions near PGMW 3204/3209, PGMW 3120, PGMW 3070, and also
possibly around PGMW 3160.
\label{n11b}}
\end{figure}

\begin{figure}
\begin{center}
\includegraphics[width=8cm]{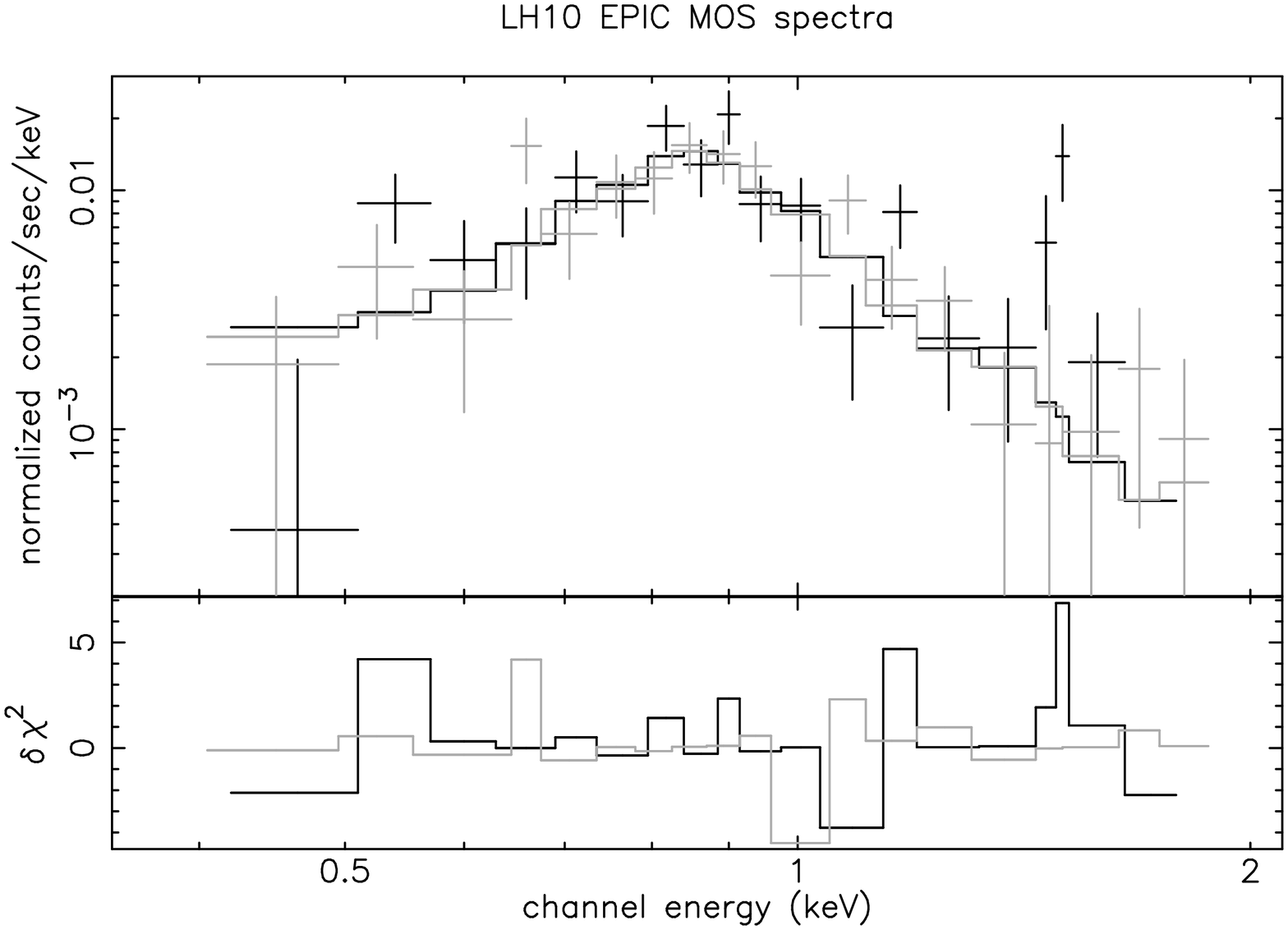}
\end{center}
\caption{The X-ray spectrum of the diffuse emission 
associated with LH10, shown along with the best fit absorbed 
$mekal$ model.  EPIC MOS1 are shown
in black and EPIC MOS2 in light grey.
\label{speclh10}}
\end{figure}

In this field, three X-ray sources have been detected: 
XMMU J045643.0$-$662504, 045646.1$-$662447, and 045648.3$-$662400 (see Fig. 
\ref{n11b}). The brightest one corresponds to the area near 
PGMW 3070 (O6V), a second source is close to PGMW 3120 (O5.5((f*))\,) 
and a last one is located to the north of these stars but is without any 
bright counterpart. Very faint diffuse emission is also detected near 
the optical pair PGMW 3209 (O3III(f*)+OB) and PGMW 3204 (O6-7V). 
All these correlate well with the positions
of wind-blown bubbles found by Naz\'e et al. (\cite{naz}). Soft, extended 
emission is also present all over the cluster (see Fig. \ref{color}).

To investigate deeper the origin of these X-ray emissions, we focused on a
region of radius 13\arcsec\ centered on PGMW 3070. Using the models derived 
for the whole region encompassing LH10 (see Table \ref{specdif} and Fig. 
\ref{speclh10}), we found that this region possesses an unabsorbed 
luminosity of $\sim3.3-3.9\times$10$^{33}$ erg 
s$^{-1}$ in the $0.1-2.$ keV range. On the other hand, following the same 
procedure as for LH9, we predicted an unabsorbed X-ray luminosity of $\sim$ 
8$\times$10$^{32}$ erg s$^{-1}$ for all stars with known spectral types in 
this small region (PGMW 3058, 3070, 3090, and 3073), i.e. a factor 4 below
the observed value. For the whole LH10 (i.e. the same region as that
shown in Fig. \ref{dif}), the emission predicted is 
5$\times$10$^{33}$ erg s$^{-1}$, while the observed value is 
$\sim2.2-2.6\times$10$^{34}$ erg s$^{-1}$. The discrepancy is thus similar.
 
Without doubt, there is a source of X-ray emission in addition to the 
massive stars themselves. Unlike LH9, hidden SNRs are very unlikely to 
contribute, since LH10 is much younger and even its most massive stars 
have not yet evolved. In addition, the youth of LH10 implies that 
most PMS stars are still in their protostellar stages, where their X-ray 
emission is strongly attenuated by their parental molecular cloud.
However, we note that the X-ray peaks are in regions where bubble expansion 
and some high-velocity motions have been discovered (Rosado et al. \cite{ros}, 
Naz\'e et al. \cite{naz}): hot shocked gas may thus be an additional source of 
X-rays. But the peak emission also appears harder than the surrounding soft 
emissions (see Fig. \ref{color}). While there is certainly soft, extended 
emission in LH10 - maybe associated with shocked winds inside a wind-blown 
bubble -, this is unlikely the only source of X-rays. A cluster wind 
and colliding 
winds in binaries may also be partly responsible for the enhancement of 
the emission.  The contributions of these different sources of 
X-rays should be disentangled by future X-ray observations, ideally done with 
a high resolution and a high sensitivity, altogether.

\section{LH 13}

LH13 is situated to the east of LH9 and LH10. It is embedded in N11C, 
one of the bright nebular components of \n, and consists of two compact 
stellar clusters, Sk $-66^{\circ}$41 and HNT, and several field stars. 
The ages of Sk $-66^{\circ}$41 and HNT have been estimated to 
$\le5$\,Myr and 100\,Myr, 
respectively, suggesting a non-association between the two clusters 
(Heydari-Malayeri et al. \cite{hey00}). The same authors proposed that the 
O stars surrounding Sk $-66^{\circ}$41 may have been ejected from the cluster 
by dynamical interactions. 

\begin{figure}
\begin{center}
\includegraphics[width=8cm]{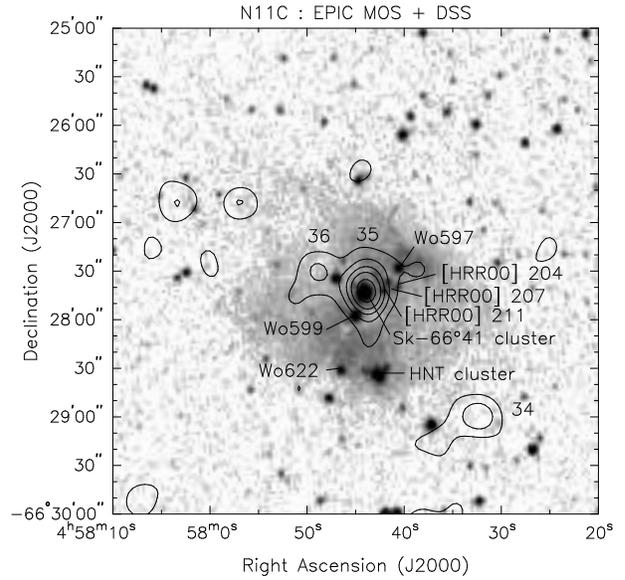}
\end{center}
\caption{ Close-up on the LH\,13 cluster embedded in the N11C nebula
(same contours as in Fig. \ref{n11l}). X-ray point sources present in this
field are labeled. The main clusters and early-type stars (Heydari-Malayeri et al. \cite{hey00}) are also shown.
\label{lh13}}
\end{figure}

\begin{figure}
\begin{center}
\includegraphics[width=8cm]{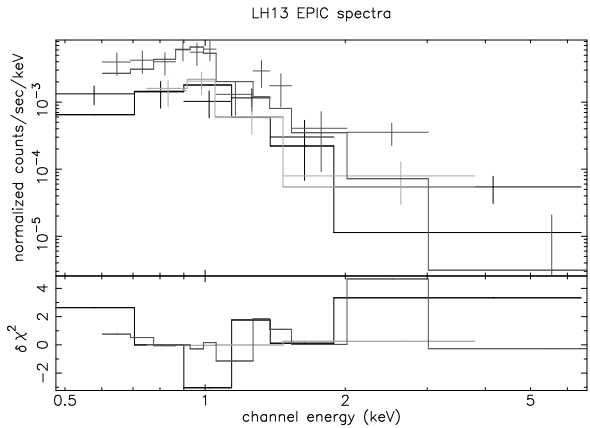}
\end{center}
\caption{The X-ray spectrum of source \#35, which corresponds to
Sk $-66^{\circ}$41, shown along with the best fit absorbed {\em mekal} 
model.  EPIC pn data are shown in dark grey, EPIC MOS1
in black and EPIC MOS2 in light grey.
\label{speclh13}}
\end{figure}

Two X-ray sources are detected in LH13. The brightest one, XMMU 
J045744.1$-$662741 (source \# 35) corresponds to Sk~$-66^{\circ}$41, 
while a fainter source, XMMU J045749.2$-$662726 (source \# 36), does 
not seem to have any obvious counterpart (see Fig. \ref{lh13}). The HNT 
cluster is not detected: this is not 
totally surprising since the brightest stars in HNT are of spectral 
type later than A0 and they should thus emit less X-rays (if any). 
Predicting the X-ray luminosity of Sk $-66^{\circ}$41 is not easy, since 
only its global properties are known: spectral type (O3V ((f*))+OB, 
Heydari-Malayeri et al. \cite{hey00}), V magnitude (11.72~mag, Heydari-Malayeri 
et al. \cite{hey87}), extinction (E(B$-$V)=$0.11-0.22$~mag, Caplan \& Deharveng 
\cite{cap85}, Heydari-Malayeri et al. \cite{hey87}). Using the same relations 
as before, we predict an unabsorbed luminosity up to 3$\times$10$^{33}$ 
erg s$^{-1}$ in the $0.1-2.$ keV energy range. 
This is slightly lower than the observed 
value $L_X^{unabs}\sim7.5\times10^{33}$ erg s$^{-1}$ for the $mekal$ models.
The situation here is thus very different from LH9 since no large 
excess emission is detected.  However, we note that the spectrum\footnote{To 
avoid contamination from the nearby XMMU J045749.2$-$662726 (source \#36), 
we had to extract the spectra in a small region some 17\farcs5 in radius. 
A correction to the encircled energy loss is made via the option 
$modelee=yes$ of the {\it arfgen} task.} of this 
source is very noisy (see Fig. \ref{speclh13}), and that a deeper X-ray
observation is needed in order to pinpoint the physical properties of 
the source. A detailed spectroscopic investigation of the content
of the cluster is also crucial to better estimate the expected X-ray
emission. 

It may seem quite strange that in the dense medium of N11C, the powerful 
massive stars in Sk $-66^{\circ}$41 have not blown any bubble. The 
observations of Rosado et al. (\cite{ros}) do not show any expansion 
with $v>10$ \kms\ or expanding regions of sizes $> 2.2$ pc. Using the Weaver 
et al. (\cite{wea}) model, a density of 20 cm$^{-3}$ (Heydari-Malayeri et 
al. \cite{hey87}) and typical parameters of an O3V star, this lack of 
detection implies that either the bubble has $R=37$~pc, corresponding 
to about 2.5\arcmin, and $t=2.2$ Myr 
(so that $v=10$ \kms) or that the dynamical timescale $t$ is unrealistically 
short ($t<<1$ Myr) so that the bubble's radius has not reached 1.1~pc 
yet. Using the former and more realistic physical parameters, we can
determine the expected X-ray luminosity of the hypothetical bubble, and we find 
$\sim$4$\times$10$^{35}$ erg s$^{-1}$ (Chu et al. \cite{chu}), which is 
clearly ruled out by our observations: the modest X-ray luminosity of LH 13
is thus most probably due to the stars of Sk $-66^{\circ}$41 alone.
However, we may note that the extinction in N11C is not uniform, and may
be rather large in some regions: Heydari-Malayeri et al. (\cite{hey87})
have actually detected extinctions E(B$-$V) up to 0.9~mag. Extended 
regions of hot gas with modest X-ray emission might thus be present 
in N11C, but the large extinction would prevent us from detecting them in our 
observation.

\section{Conclusion}

We report in this paper the observations by the {\it XMM-Newton} 
satellite of the giant \hii\ region \n\ situated in the LMC. 
In this field, we have detected large areas of soft diffuse X-ray emission 
and 37 point sources, one of which is apparently associated with a 
small and poorly known extended object, BSDL 188. 

\n\ harbours a wealth of phenomena associated with massive stars: it
contains a SNR, N11L, and four OB associations at different stages 
of evolution and interaction with their surroundings (LH9, LH10, LH13 
and LH14). All are detected in the {\it XMM-Newton} data, except for 
LH14, which is unfortunately outside of the \x\ FOV. The stars from LH9, 
the largest cluster of \n, have blown a superbubble which is detected 
in X-rays as a large soft X-ray emission. It peaks near HD 32228, 
a dense cluster containing a WR star and several OB stars,
and it is rather well confined within the \ha\ filaments delineating 
the superbubble. The combined emission of all individual massive stars 
of LH9 can not explain the high level of emission observed, nor can the
contribution from the hot shocked gas of the superbubble as predicted 
by the Weaver et al. (\cite{wea}) model. Hidden SNRs, colliding-wind 
binaries and the often neglected pre-main sequence stars most 
probably provide additional X-ray emission. It is probably a combination
of all these effects that will resolve the discrepancy, since an estimation
of the contribution of the latter alone has been shown insufficient
to explain the whole X-ray emission. Moreover, we do not exclude 
that the superbubble is leaking some hot gas since faint, soft emission 
is detected to the south of the cluster. 

The action of LH9 on its 
surroundings has probably triggered the formation of a second cluster, 
LH10. The X-ray emission from this younger cluster consists of three 
rather pointlike sources, in addition to a soft extended emission of 
reduced intensity. The two brightest sources are not centered on the 
most powerful stars of the cluster, i.e. the young O3 stars, but 
seem associated with the expanding bubbles blown by stars in the SW 
part of LH10 (Naz\'e et al. \cite{naz}). The total X-ray emission of 
LH10 is four times larger than the total emission from its stellar 
components. Since the cluster is still very young, the excess emission 
can probably not be attributed to hidden SNRs or active T Tauri stars, but 
it probably originates from the hot gas filling the wind-blown bubbles. 
To completely understand the X-ray emission from these two clusters, 
disentangling the individual contributions (extended sources vs. 
a simple accumulation of non-resolved point sources) is of the utmost 
importance. However, this is not feasible in a reasonable amount of 
time with the current X-ray observatories, but \n\ will be a perfect 
target for the next generation of X-ray satellites.

In contrast, the X-ray emission from LH13 does not show any extended 
emission and could be well explained by the stars of the Sk $-66^{\circ}$41 cluster 
alone. To the west of the field, the SNR N11L shows hot gas outside 
the optical bubble, which is associated with an additional radio emission.
The spectra of the SNR and of the extended X-ray plume to its north are 
similar, 
though with a larger absorbing column for the latter. For the first time, 
the SNR was also detected in the ultraviolet, and it presents at these
wavelengths a morphology very similar to the optical one.

Note that the high excitation blob N11A is not detected in our 
observation, but this is not completely surprising since at least
the theoretically expected X-ray flux from the stellar population 
embedded in this nebula (Heydari-Malayeri et al. \cite{hey01}) is 
very small. 

During our X-ray observation, the OM camera onboard {\it XMM-Newton}
has provided unique UV photometry of more than 6000 sources. This 
photometry is available to the scientific community through CDS.

\begin{acknowledgements}
Y.N. thanks Micha\"el De Becker and Hugues Sana for their help and 
valuable discussion on the sources' variability, and also Dr H. Dickel
for a helpful discussion about CO observations.
I.I.A. acknowledges support from the Russian Foundation for Basic 
Research through the grant No 02-02-17524.
We acknowledge support from the PRODEX XMM-OM and Integral Projects 
and through contracts P4/05 and P5/36 `P\^ole d'Attraction Interuniversitaire' 
(Belgium). This publication makes use of data products from USNO B1.0, the Two Micron All Sky Survey and the Guide Star Catalog-II. 
The authors are grateful to an anonymous referee for valuable comments 
that helped to improve our manuscript.

\end{acknowledgements}

\end{document}